\documentclass[english,twocolumn]{article}
\usepackage{amsmath}
\usepackage{amssymb}
\usepackage{cite}

\makeatletter

\usepackage{babel}

\begin{document}
\begin{titlepage}
\thispagestyle{empty}

\bigskip

\begin{center}
\noindent{\Large \textbf
{Hidden Conformal Symmetry in Randall-Sundrum 2 Model: Universal Fermion Localization by Torsion}}\\

\vspace{0,5cm}

\noindent{G. Alencar\footnote{e-mail: geova@fisica.ufc.br }}

\vspace{0,5cm}

 {\it Departamento de F\'{\i}sica, Universidade Federal do Cear\'{a}-
Caixa Postal 6030, Campus do Pici, 60455-760, Fortaleza, Cear\'{a}, Brazil. 
 }

\end{center}

\vspace{0.3cm}

\begin{abstract}
In this manuscript we describe a hidden conformal symmetry of the
second Randall-Sundrum model (RS2). We show how this can be used
to localize fermions of both chiralities. The conformal symmetry leaves few free dimensionless constants
and constrains the allowed interactions.
In this formulation the warping of the extra dimension emerges from a partial breaking
of the conformal symmetry in five dimensions. The solution of the
system can be described in two alternative gauges: by the metric or
by the conformon. By considering this as a fundamental symmetry
we construct a conformally invariant action for a vector
field which provides a massless photon localized over a Minkowski brane. This
is obtained by a conformal non-minimal coupling that breaks the gauge symmetry
in five dimensions. We further consider a generalization of the model
by including conformally invariant torsion. By coupling torsion non-minimally
to fermions we obtain a localized zero mode of both chiralities completing
the consistence of the model. The inclusion of torsion introduces a fermion 
quartic interaction that can be used to probe the existence of large extra dimensions
and the validity of the model. This seems to point to the fact that
conformal symmetry may be more fundamental than gauge symmetry and that
this is the missing ingredient for the full consistence of RS scenarios. 
\end{abstract}
\end{titlepage}

In the beginning of the $2000$ some models of large extra dimensions
emerged\cite{ArkaniHamed:1998rs,Randall:1999vf}. Differently of the Kaluza-Klein mechanism, in which
a compact extra dimension is introduced, in these models some mechanism
must confine the fields to the brane, a $(3+1)-$subspace of a higher
dimensional theory. In Ref. \cite{ArkaniHamed:1998rs} a mechanism has been found that
trap gauge fields, however it does not work for gravity. Soon after,
a solution to the confining of gravity was found by the introduction
of a non-factorizable metric\cite{Randall:1999vf}. In this model, called Randall-Sundrum
2 (RS2), gravity is confined to the brane. However soon became clear
that other fields are not localized over the brane. In fact only
gravity and scalar fields are trapped and this became a central problem
in models of large extra dimensions, since for example fields of spin one neither $1/2$ are trapped\cite{Bajc:1999mh}. 

In this manuscript we present a model based on a hidden conformal symmetry of the RS2 model that 
solves the problems mentioned in the previous paragraph.  That is, our model confines to the brane:
gravity, vector fields, and, importantly, fields of spin $1/2$
for positive or negative tension branes, a long standing problem in RS2 models. 

Consider the $D-$dimensional action defined by the Lagrangian
\begin{equation}\label{scalargravity}
L_{G}=\xi\chi^{2}R-\frac{1}{2}\chi\nabla_{M}\partial^{M}\chi-U(\chi)
\end{equation}
and the conformal transformation $\tilde{g}{}_{MN}=e^{2\rho}g_{MN}$. 
The determinant transforms as $\sqrt{-\tilde{g}}=e^{D\rho}\sqrt{-g}$ and therefore has weight $D$. With this an conformally invariant action is obtained if the
Lagrangian has weight $-D$. This fixes the transformation of the scalar field to be $\tilde{\chi}=e^{-\frac{D-2}{2}\rho}\chi$ and $U(\chi)=u_0\chi^{n}$ with $n=2D/(D-2)$ with $u_0$ a free parameter. In order to cancel the terms with derivatives in $\rho$ we also need that $8\xi=(D-2)/(D-1)$ and $\chi$ is called conformon field \cite{Faraoni:1998qx}. From now on we must consider $D=5$ and $z$ the extra dimension coordinate. By partially breaking the conformal symmetry
we can choose $\chi=\chi(z)$. With this the action keeps a residual
conformal symmetry with $\rho=\rho(z)$,
\begin{equation}
L_{G}=\xi\chi^{2}R-\frac{1}{2}\chi(z)\nabla^2\chi(z)-U(\chi(z)),
\end{equation}
where the prime means a $z$ derivative. At this point a $(D-1)-$brane Lagrangian with tension $\lambda(\chi)$ can be introduced. In order to preserve the residual transformation 
the brane term must be coupled to the conformon giving us $\lambda(\chi)=\mu\chi^{n}(z)\delta(z)$, with $\mu$ a second free parameter. Note that, $\lambda(\chi)=\mu\chi^{10/3}(z)\delta(z)$ and $U(\chi)=u_0\chi^{10/3}$
are imposed to preserve the residual
conformal transformation. As we will see soon this term can be obtained
by localizing zero modes off matter fields. We then get for the full Lagrangian of the model 
\begin{equation}\label{conformalRS2}
L=\xi\chi^{2}R-\frac{1}{2}\chi(z)\nabla^2\chi(z)-U(\chi)-\lambda(\chi)\delta(z).
\end{equation}
The residual symmetry can be used to describe the system in two different
ways. The first is by fixing the conformon field to a constant $\chi=\chi_{0}$ 
\begin{equation}
L=\xi\chi_{0}^{2}R-u_0\chi_{0}^{n}-\mu\chi_{0}^{m}\delta(z)
\end{equation}
and we see that breaking the residual conformal symmetry is the same as fixing
an energy scale for gravity by choosing $\xi\chi_{0}^{2}=2M^{3}$.
With this we obtain the full RS2 model in the standard form. We are
left with the free parameters $u_0$ and $\mu$. The first can provide the cosmological
constant if we choose $u_0\chi_{0}^{n}=\Lambda$. One solution
to this is given by $ds^{2}=e^{2A(z)}\eta_{MN}dx^{M}dx^{N}$ with
\begin{equation}
A(z)=-\ln(k|z|+1);\mu=24M^{3}k;\Lambda=-24M^{3}k^{2}.
\end{equation}

The second way of describing the system is by fixing the warp factor
to $e^{A}=1$. With this we get the Lagrangian
\begin{equation}
L=\frac{1}{2}\chi\chi''-u\chi^{n}-\mu\chi^{n}\delta(z)
\end{equation}
with equation of motion
\begin{equation}
\chi''-nu_0\chi^{n-1}-n\mu\chi^{n-1}\delta(z)=0.
\end{equation}

The solution to the above equation is $\chi=C/(k|z|+1)^{3/2}$
with $C^{4/3}=9k^{2}/8u_0$ and $\mu=4u_0/5k$. By
replacing the Minkowski metric by $g_{\mu\nu}$ the gravity part of
the action becomes
\begin{equation}
S=\xi C\int\frac{dz}{(k|z|+1)^{3}}\int d^{4}xR_{4}
\end{equation}
and in order to recover four dimensional gravity we must impose the
additional constraint $\xi C^2=2M^{3}$. With this all the constants
of the model agree in both gauges. 

Now we must construct a bulk matter action 
with a zero mode localized over the brane. We will show that imposing conformal symmetry and non-minimal couplings to gravity naturally provides a guiding mechanism for localizing other fields. Consider first the gauge field action. The construction of such an action is
problematic since the standard action, given by $F_{MN}F^{MN}$, with
$F_{MN}=\partial_{M}A_{N}-\partial_{N}A_{M}$ is conformal only in
four dimensions. In $D$ dimensions the present model provides the conformon field and
we can construct the Lagrangian $L=\chi^{2\frac{D-4}{D-2}}F^{2}$,
which has the right weight. However, as said before, we can choose
$\chi$ a constant and this is reduced to the standard Lagrangian
which does not provides a localized zero mode. To solve this problem it has been recently
shown that a non-minimal coupling to gravity can render a gauge invariant
theory in four dimensions\cite{Alencar:2014moa}. The coupling is given by $\gamma RA^{2}$, where $\gamma$ is a dimensionless coupling constant which is fixed by demanding localization of the zero mode. Despite the fact that this is not gauge
invariant, it has been shown that it generates a gauge invariant action
in a Minkowski brane. The dimensionless of $\gamma$ suggests that this term can have a conformal origin. In fact the conformal version of this term is
given by
\begin{equation}
\gamma(\xi\chi^{2}R-\frac{1}{2}\chi\nabla_{M}\partial^{M}\chi)A^{2}.
\end{equation}
We can see that by choosing the gauge $\chi=\chi_{0}$ this reduces
to the geometrical coupling used in Ref \cite{Alencar:2014moa}. Therefore the conformal
symmetry and non-minimum coupling seems to be the fundamental ingredients for obtaining a consistent
brane model. 

Now we proceed to consider the trapping of spin half fields. The free Dirac action is given by
\begin{equation}\label{Diracaction}
S=\int d^5 x \bar{\Psi}\Gamma^MD_M\Psi
\end{equation}
where $D_M=\partial_M+\omega_M$, $\omega_M$ being the spin connection. This action is conformally invariant in any $D$ with $\Psi'=e^{-(D-1)/2}\Psi$. In the above metric we get the equations of motion
\begin{equation}
\gamma^{\mu}\partial_{\mu}\Psi(x,z)+\gamma^5(\partial_z+2A')\Psi(x,z)=0.
\end{equation}
Now considering the separation of variables $\tilde{\Psi}(x)\psi(z)$ and that the zero mode satisfy the four dimensional massless equation $\gamma^{\mu}\partial_{\mu}\tilde{\Psi}(x)=0$ we obtain an effective action for this mode given by
\begin{equation}\label{effectiveDirac}
S=\int dz e^{4A}\psi^2(z)\int d^4 x \bar{\tilde{\Psi}}(x)\gamma^{\mu}\partial_{\mu}\tilde{\Psi}(x)
\end{equation}
with $\psi(z)$ determined by the equation
\begin{equation}\label{zeromode}
\psi'+2A'\psi=0.
\end{equation}
The solution to the above equation is $\psi(z)\propto e^{-2A}$ and with this the effective action (\ref{effectiveDirac}) is not finite.  By introducing an Yukawa interaction with a scalar
field the localization of one of the chiralities can be obtained\cite{Kehagias:2000au}. However up to now no mechanism has been found that fully localize both chiralities. 
Despite the fact that the Dirac action is conformal  in any dimensions it does not presents a
localized zero mode. From the equation (\ref{zeromode}) we see that a simple solution would be to change the coefficient of $A'$ and we must couple the Dirac field to a first derivative of the metric. The only possibility in this context is the spin connection since the Ricci scalar has second derivatives and another kind of non-minimal coupling between gravity and fermions must be added. The missing ingredient
is a conformal coupling of the fermion field with torsion. For this we must
generalize our previous result to introduce a conformally invariant
torsion theory˜\cite{Obukhov:1982zn}. Torsion can be introduced
by adding an antisymmetric part to the standard affine connection $\tilde{\Gamma}_{\quad NP}^{M}=\Gamma_{\quad NP}^{M}+K_{\quad NP}^{M}$,
where $\Gamma_{\quad NP}^{M}$ is the symmetric affine connection
and $K_{\quad NP}^{M}$ is called the contorsion tensor. We also define
$2K_{\quad NP}^{M}=T_{\quad NP}^{M}-T_{N\;P}^{\;M}-T_{P\;N}^{\;M}$.
The action invariant under conformal transformation is given by\cite{Obukhov:1982zn,Shapiro:2001rz}
\begin{equation}\label{torsionaction}
S=\xi\int d^{5}x\chi^{2}(R-4\nabla_{M}T^{M}-3T^{2}+\frac{1}{2}q_{MNP}^{2}+\frac{1}{24}S_{MN}^{2})
\end{equation}
where $R$ and $\nabla_{M}$ are the Ricci scalar and the covariant derivative computed from $\Gamma_{\quad NP}^{M}$. We also define $T_{M}\equiv T_{\quad MP}^{P}$ as the trace, $S^{MN}\equiv\varepsilon^{MNOPQ}T_{OPQ}$ the antisymmetric and $q_{MNP}$ the traceless decomposition satisfying $\varepsilon^{MNOPQ}q_{OPQ}=0$. The quantity between parenthesis in Eq. (\ref{torsionaction}) is obtained by computing the Ricci scalar $\tilde{R}$ from $\tilde{\Gamma}_{\quad NP}^{M}$. The model is constructed such that derivatives in $\rho$ cancel in $\tilde{\Gamma}_{\quad NP}^{M}$ and this guarantee that (\ref{torsionaction}) is invariant. With this we also get that $S_{MN}$ and $q_{MNP}$ are conformally invariant and $T'_{M}=T_{M}-2\partial_{M}\rho$ with the other fields transforming as before. If we solve the equations of motion for torsion we see
that $q_{MNP}$ and $S_{MN}$ do not posses dynamics and can be eliminated
from the action˜\cite{Obukhov:1982zn}. The equation of motion for $T_{M}$ is given by $3T_{M}=4\partial_{M}\ln\chi$ that also can be eliminated from (\ref{torsionaction}) and we recover the torsionless conformal action (\ref{scalargravity}). Note that if the above
model is coupled to fermions, the minimal coupling preserving conformal
symmetry is given by
\begin{equation}\label{minimalDirac}
S=\int d^{5}x\bar{\Psi}(\Gamma^{M}\nabla_{M}-\frac{1}{96}\tilde{\Gamma}_{MN}S^{MN})\Psi
\end{equation}
where $\tilde{\Gamma}_{MN}\equiv\varepsilon_{MNPQR}\Gamma^{PQR}$
and now we see that by evaluating the equation for $S_{MN}$ we get
\begin{equation}\label{EqS}
S_{MN}=\frac{1}{4\chi^{2}}\bar{\Psi}\tilde{\Gamma}_{MN}\Psi
\end{equation}
and this will generate a quartic interaction term in the action (\ref{torsionaction}).
This term alone is not enough to localize the field. In fact, just
as in the case of the gauge field, we must look for a non-minimal
coupling of the fermion field with torsion. Note that now we have
a new compensating field $T_{M}$, with transformation depending on
the first derivative of $\rho$. With this, a conformal invariant non-minimal
coupling is given by
\begin{equation}
\theta\bar{\Psi}\gamma^{M}(T_{M}-\frac{4}{3}\partial_{M}\ln\chi)\Psi
\end{equation}
and the equations of motion for the fermion field is given by
\begin{equation}
[\Gamma^{M}\nabla_{M}-\frac{1}{96}\tilde{\gamma}_{MN}S^{MN}+\theta\gamma^{M}(T_{M}-\frac{4}{3}\partial_{M}\ln\chi)]\Psi=0.
\end{equation}

Now, in order to solve for the localization of the fermion field,
we choose the gauge $\chi=\chi_{0}$ with the metric given by $g_{MN}=e^{2A}\eta_{MN}$.
The way to compute this is simplified if we use the conformal transformation
of $T_{M}$ and we get $T_{M}=-2\partial_{M}A$. Since $S_{MN}$ is
invariant for conformal transformations it is zero for the above metric
and we get the simple equation
\begin{equation}
\Gamma^{M}(\nabla_{M}-\frac{4}{3}\theta\partial_{M}A]\Psi=0.
\end{equation}

With this and by performing the standard manipulations we get for
the zero mode equation
\begin{equation}
\psi'+2(1-\frac{2}{3}\theta)A'\psi=0
\end{equation}
with normalized solution $\psi=\sqrt{(5k\theta/3)}e^{-2(1-2\theta/3)A}$ and the condition
for localization is given by $\theta>3/8$. In the case of the vector field the dimensionless parameter of the non-minimal coupling is fixed by demanding the localization of the field\cite{Alencar:2014moa}. With this we latter have shown that  a precise phenomenological prevision of the model can be obtained\cite{Alencar:2015rtc}. However the fact that $\theta$ is a free parameter subjected to the above simple constraint may becomes an advantage of the model. As said before, upon substituting (\ref{EqS}) in (\ref{torsionaction}) we can get a quartic interaction and using the above solution we get that the coupling is proportional to $~\theta M^{-2}_{pl}$. Despite the fact that $M^{-2}_{pl}\sim10^{-38}GeV^{-2}$ is very small this new free parameter can be fixed by possible showing up of torsion in phenomenological data without the necessity for additional new dynamics or mass scales\cite{Belyaev:1998ax}. Therefore torsion is necessary in order to have a consistent and well defined four dimensional brane in warped scenarios with one large extra dimension. It is important to further note  that the above solution to the normalized zero mode is valid for any warp factor $A(z)$ and all the above results are valid for any smooth versions of the RS2 model. This kind of property is also present in the case of vector field localization and seems to be an universal aspect of our model. However we should point that here we have considered only zero modes of fermion fields. Massive fields are important since in the standard model fermions are massive. Issues related to this and to resonances of fermions will be considered in the near future. We also point that by demanding conformal symmetry we obtain that the brane term in the action (\ref{conformalRS2}) must be coupled to the conformon field. This kind of coupling is necessary for getting a stabilized radius in the compact RS1 model\cite{DeWolfe:1999cp}. The radius of the extra dimension will be related to a small parameter introduced to break the conformal symmetry \cite{geova}. The possibility of smooth version of the present model has also been proposed recently by us\cite{Alencar:2017vqd}.  Therefore it seem that a conformal version of RS1 may provides more natural solution to the this problem. All this points to the fact that the above model is both theoretically and phenomenologically interesting. 

\section*{Acknowledgments}

We would like to thanks A. A. Moreira for discussions and motivations. We also acknowledge the financial support provided by Funda\c c\~ao Cearense de Apoio ao Desenvolvimento Cient\'\i fico e Tecnol\'ogico (FUNCAP), the Conselho Nacional de 
Desenvolvimento Cient\'\i fico e Tecnol\'ogico (CNPq).

\end{document}